\documentclass[sigconf]{acmart}
\AtBeginDocument{%
  \providecommand\BibTeX{{%
    \normalfont B\kern-0.5em{\scshape i\kern-0.25em b}\kern-0.8em\TeX}}}

\setcopyright{acmlicensed}
\copyrightyear{2018}
\acmYear{2018}
\acmDOI{XXXXXXX.XXXXXXX}

\acmConference[Conference acronym 'XX]{Make sure to enter the correct
  conference title from your rights confirmation emai}{June 03--05,
  2018}{Woodstock, NY}
\acmISBN{978-1-4503-XXXX-X/18/06}

\usepackage{amsmath}

\usepackage{subfigure}
\usepackage{graphicx}
\usepackage{soul}
\usepackage{subcaption}
\usepackage{multirow}
\usepackage{multicol}
\begin{document}


\title{Incorporating Group Prior into Variational Inference for Tail-User Behavior Modeling in CTR Prediction}
% Enhancing Tail User Preferences in Click-Through Rate Prediction with Group Prior Variational Inference and Volume-Preserving Flows

\author{Han Xu}
\affiliation{%
  \institution{Kuaishou Technology}
  \city{Beijing}
  \country{China}
}
\email{xuhan03@kuaishou.com}

\author{Taoxing Pan}
\affiliation{%
  \institution{Kuaishou Technology}
  \city{Beijing}
  \country{China}
}
\email{pantaoxing@kuaishou.com}

\author{Zhiqiang Liu}
\affiliation{%
  \institution{Kuaishou Technology}
  \city{Beijing}
  \country{China}
}
\email{zhiqliu1103@gmail.com}

\author{Xiaoxiao Xu}
\affiliation{%
  \institution{Kuaishou Technology}
  \city{Beijing}
  \country{China}
}
\email{xuxiaoxiao05@kuaishou.com}

\author{Lantao Hu}
\affiliation{%
  \institution{Kuaishou Technology}
  \city{Beijing}
  \country{China}
}
\email{hulantao@kuaishou.com}

\renewcommand{\shortauthors}{Trovato and Tobin, et al.}

\begin{abstract}

User behavior modeling---which aims to extract user interests from behavioral data---has shown great power in Click-through rate (CTR) prediction, a key component in recommendation systems. Recently, attention-based algorithms have become a promising direction, as attention mechanisms emphasize the relevant interactions from rich behaviors. However, the methods struggle to capture the preferences of tail users with sparse interaction histories. To address the problem, we propose a novel variational inference approach, namely \textbf{G}roup \textbf{P}rior \textbf{S}ampler \textbf{V}ariational \textbf{I}nference (GPSVI), which introduces group preferences as priors to refine latent user interests for tail users. In GPSVI, the extent of adjustments depends on the estimated uncertainty of individual preference modeling. In addition, We further enhance the expressive power of variational inference by a volume-preserving flow. An appealing property of the GPSVI method is its ability to revert to traditional attention for head users with rich behavioral data while consistently enhancing performance for long-tail users with sparse behaviors. Rigorous analysis and extensive experiments demonstrate that GPSVI consistently improves the performance of tail users. Moreover, online A/B testing on a large-scale real-world recommender system further confirms the effectiveness of our proposed approach.

\end{abstract}





\received{20 February 2007}
\received[revised]{12 March 2009}
\received[accepted]{5 June 2009}

\maketitle

\section{Introduction}
% introduction 
% 先说 CTR 预估任务很重要，然后说 user behavior model 很重要
% 然后说常见的 attention-based method 在长尾用户上容易碰到问题（这里要补充related work)

Click-through rate (CTR) prediction is a fundamental task for many large-scale applications, such as e-commerce and short-video sharing platforms \cite{kalman,constrained-rl-kuaishou}. Regarding each user's behaviors as a sequence in chronological order, extracting user preferences behind given historical behaviors has achieved great improvement \cite{DIN,DIEN,SIM,CIN}. Consequently, modeling users' complex sequential behaviors is a key component in CTR prediction.

Initially, researchers focus on learning better representation of user historical behaviors \cite{DIN,DIEN,DSIN}. For example, DIN \cite{DIN} introduces target attention, which emphasizes target-relevant behaviors and suppresses irrelevant ones. In \cite{DIEN}, DIEN proposes an interest-evolving layer to capture the dynamic evolution of user interests about the target items. Recent works endeavor to leverage rich user behaviors under the constraints of simple model deployment and system latency, such as MIMN \cite{MIMN} and SIM \cite{SIM}. These popular approaches achieve promising performance when most users' historical behavioral data is sufficient enough because they assume that user interests are included within their behaviors. Consequently, the previous user behavior modeling methods are sensitive to the scale and quality of historical behavior data. However, in practical applications, the number of user behaviors inherently follows a long-tail distribution \cite{learningtrans}. Sequence lengths for head users can range from hundreds to thousands, while tail users have only a few or even none. Notably, due to the limited opportunities for feedback from long-tail users, their historical behaviors alone are insufficient for representing their interests. Existing deterministic methods primarily focus on the historical behavior of tail users, leading to inaccurate interest modeling and thereby resulting in misguidance in the CTR prediction system. Our experiments further reveal that while attention-based models yield promising results with behavior-rich head users, they often overlook many behavior-poor tail users and degrade the performance of CTR tasks.

% However, existing user behaviors modeling methods struggle to precisely capture long-tail users preference from their short behaviors sequences. Since short behavior sequences means lack of user interaction records, user interest extracted from these sequences is incomplete and unreliable. In such case, existing deterministic behavior modeling modules will mislead the CTR prediction. To tackle this problem, we propose a novel Group Prior Sampler Variational Inference (GPSVI) approach based on the uncertainty estimation. Considering the group preference as a proper prior for long-tail users, the proposed method incorporate the behavior modeling with the prior for correction, and the method adaptively controls the magnitude of correction by the confidence of the user interest modeling. Consequently, an appealing property of GIVM method is that it can degenerate to traditional attention for high active users and achieve the consistent improvement for long-tail users. 

In recent years, a number of studies aim to enhance tail users' experiences by improving embedding generalization ability, such as content-based methods \cite{image-cold-start,collective-cold,NIPS2017_dbd22ba3}, meta-learning involved methods \cite{warm-up,sequential-meta}, and variational inference methods \cite{velf}. Although these works make some progress, capturing genuine tail users' interests remains an unresolved challenge. To address the problem, we propose a novel Group Prior Sampler Variational Inference (GPSVI) approach based on uncertainty estimation. Regarding group preferences as an appropriate prior for long-tail users, the proposed method integrates behavior modeling with this prior for correction. Additionally, the method adaptively adjusts the extent of correction based on the confidence level in the user interest modeling. We further enhance the expressive power of variation inference by adding a normalizing flow. An attractive property of the GPSVI method is its ability to revert to traditional attention mechanisms for head users while consistently enhancing performance for long-tail users.

We conduct extensive experiments and rigorous analysis to verify the effectiveness and generality of our proposed method. Our experiments are conducted in a large-scale industry dataset collected from a real-world application and a famous public Amazon dataset \cite{mcauley2015amazon}. The experiments indicate that GPSVI consistently and significantly outperforms the state-of-the-art user behavior modeling methods Trans \cite{Trans}. Rigorous ablation studies demonstrate the necessity of our proposed modules, including the group prior based sampler and the volume-preserving flow. Furthermore, we test our proposed algorithm through online experiments and observe a great improvement. Notice that unlike head users, who already present satisfaction with little scope for further enhancement, improving the performance of tail users offers a greater opportunity for significant gains. Compared to the baseline model, our method results in a 0.306\% increase in overall CTR, with a specific improvement of 0.659\% among tail users. Our methods have been successfully deployed to serve the main traffic for two months.

\section{Methods}
CTR prediction is a critical task in industry applications,  such as e-commerce search engines and recommendation systems. Given a user, a candidate item, and the contexts in an impression scenario, CTR prediction is used to infer the probability of a click event. The mainstream methods for click-through rate (CTR) prediction mostly adopt the Embedding\&Network paradigm, which takes several critical fields of features as input, such as user profile, item profile, context and user behaviors. Denote the features mentioned above as $u, i, c, s$. Then, the output of the CTR model is 
$$\hat{y} = f(u,i,c,s),$$
where $\hat{y}$ is the prediction. 

Among them, user behaviors are records of users' interactions on specific items, faithfully reflecting users’ immediate and evolving interests. Consequently, recent works consider these features as a key for user interest modeling, and propose extensive deterministic attention-based user behavior modeling approaches. They aim to extract the user's preference $\hat{\mathbf{v}}_q$, when given the specific item $\mathbf{q} \in \mathbb{R}^d$ and the historical behavior sequence with length $L$. We denote the key vector of the $l$-th behavior as $\mathbf{k}_l \in \mathbb{R}^d$, and the corresponding value vector as $\mathbf{v}_l \in \mathbb{R}^d$. The keys and values are packed together into matrices $K$ and $V$, i.e.,
 \begin{align*}
    K = (\mathbf{k}_1, \mathbf{k}_2, \cdots, \mathbf{k}_L)^\top,\\
    V = (\mathbf{v}_1, \mathbf{v}_2, \cdots, \mathbf{v}_L)^\top.
\end{align*}
The commonly seen attention mechanism for user behavior modeling is as follows, i.e.,
\begin{align*}
    \alpha_l = \frac{\exp\left(\mathbf{q} \mathbf{k}_l^\top\right)}{\sum_{\tau=1}^L \exp\left(\mathbf{q} \mathbf{k}_\tau^\top\right)}, \quad \hat{\mathbf{v}}_q = \sum_{l=1}^L \alpha_l \mathbf{v}_l.
\end{align*}

However, existing user behaviors modeling methods struggle to precisely capture long-tail users preferences from their scarce behaviors. Since short behavior sequences indicates a lack of user interaction records, user interest extracted from these sequences may be incomplete and unreliable. In such cases, existing deterministic behaviour modelling modules will mislead the CTR prediction. To tackle this problem, we propose a novel Group Prior Sampler Variational Inference (GPSVI) approach based on uncertainty estimation. Considering the group preference as a proper prior for long-tail users, the proposed method incorporates the behavior modeling with the prior for correction, and the method adaptively controls the magnitude of correction by the confidence of the user interest modeling. Consequently, an appealing property of the GPSVI method is that it can degenerate to traditional attention for highly active users and achieve consistent improvement for long-tail users.

% 先介绍方法分成哪几个模块，然后说明每个模块各自是什么功能
% 分哪几个模块呢？引入群体先验的 VAE 算一个，volumn preserving flow 也算一个。

\subsection{Variational Framework with Group Prior Sampler} \label{section:Group_Prior}
% 先介绍 VI，然后再介绍 group prior sampler
% VI 需要介绍什么？
% 1, 先说我们把行为序列建模输出看成是一个分布，然后引入VI，简单介绍 ELBO
% 2, 然后介绍 group prior sampler，说清楚群体先验是如何引入的，然后在这种情况下修正 重参数 以及 ELBO
Given the output of the attention-based behavior modeling module as input, the GPSVI method learns the posterior distribution $q_{\phi}(\mathbf{z}|\mathbf{v})$ of latent variable $z$, where $\mathbf{z}\in \mathbb{R}^d$ and $\mathbf{v}\in \mathbb{R}^d$. Similar to general variational inference methods, GPSVI defines the posterior distribution as an isotropic Gaussians with diagonal covariance, i.e., $z\sim \mathcal{N}(\mathbf{\mu}, \text{diag}(\mathbf{\sigma}))$, where $\mathbf{\mu}\in \mathbb{R}^d$, $\mathbf{\sigma}\in \mathbb{R}^d$, and $\text{diag}(\mathbf{\sigma}$ is the diagonal matrix with diagonal entries $\mathbf{\sigma}$. Then, the decoder, parameterized by $\theta$, in GPSVI predicts the probability of click events conditional on the latent variable, that is, $\hat{y} = p_\theta(\hat{y}|\mathbf{z})$. The training objective is to maximize the Evidence Lower Bound (ELBO):
\begin{align}
    \mathcal{L}(\phi, \theta) &=ELBO(\phi, \theta) \nonumber\\
    &= \mathbb{E}_{\mathbf{z}\sim q_\phi(\mathbf{z}|\mathbf{v})}[\log p_\theta(\mathbf{y}|\mathbf{z})] + D_{KL}\left(q_\phi(\mathbf{z}|\mathbf{v})||p(\mathbf{v})\right).
\end{align}
The choice of prior and posterior distributions has a significant impact on the quality of inferences. In GPSVI, the prior is the standard normal, $p(\mathbf{v}) = \mathcal{N}(0,I)$. The prior is both appropriate and computationally convenient, before obtaining the information about the user interest \cite{Kingma2014,bowman-etal-2016-generating}. As for the posterior $q_\phi(\mathbf{z}|\mathbf{v})$, learnable parameters $\mathbf{\mu}$ and $\mathbf{\sigma}$ are derived by a neural network \cite{normalizing_flow}. For the mean $\mu$, GPSVI applies an identity transformation, i.e., $\mu = \mathbf{v}$. The identify transformation makes much sense as it maintains the concept of “attention”. GPSVI computes the $\mathbf{\sigma}$ by a neural layer with exponent activation, to ensure positive values. Since the estimation variance decreases as the number of behaviors increase, GPSVI adds a monotonic regularizer, that is 
\begin{align}\label{eq:monotonic}
    R_m = -\sum_{(u_i, u_j)}& \left(\sum_{m=0}^d \max\left(0,\sigma_{u_i}^m - \sigma_{u_j}^m\right)\mathbb{I}\left(l_{u_i} > l_{u_j}\right) \right. \nonumber \\ 
    + &\left.\max\left(0,\sigma_{u_j}^m - \sigma_{u_i}^m\right)\mathbb{I}\left(l_{u_j} > l_{u_i}\right)\right), 
\end{align}
where $l_{u_i}, l_{u_j}$ are the lengths of behavior sequences of user $u_i$ and $u_j$ respectively and $\sigma^m$ is the element in $m$-th entry of $\sigma \in \mathbb{R}^d$.

The core component of GPSVI is the group-prior-based sampler, which shifts the latent representation of the user interest by adding the group prior. Coarse features of a user $c(u)$, such as the age, the gender and the location, categorize the individual into a specific group. GPSVI models the group interest $\mathbf{g} \in \mathbb{R}^d$ for a specific item $i$ by an mlp, parameterized by $\varphi$, that is, 
\begin{align}
 g = f_{\varphi}(c(u), i) \in \mathbb{R}^d.
\end{align}
In general, the diagonal entry vector $\mathbf{\sigma}$ is considered as the uncertainty of the latent user interest. Regardless of uncertainty, deterministic methods place excessive trust in the user behavior modeling, while probabilistic methods introduce random vectors with covariance matrix $\text{diag}\mathbf{\sigma}$. Both of these methods struggle to capture genuine user interests from scarce behaviors. An interesting insight is that the group interest is an appropriate prior for the individual interest \cite{velf}. Based on the insight, the sampler in GPSVI shifts the latent variable by adding the projection of a random vector onto the plane spanned by group interest vectors. Incorporating the reparameterization technique, the sample of the latent variable $\mathbf{z}$ is 
\begin{align}
    &\mathbf{z} = \mathbf{\mu} + \text{Proj}_{\mathbf{g}}(\text{diag}(\mathbf{\sigma})\xi), \quad \xi \sim \mathcal{N}(0, \mathbf{I}), \\
    &\text{Proj}_{\mathbf{g}}(\mathbf{y})  = \frac{\langle \mathbf{y}, \mathbf{g} \rangle}{\|\mathbf{y} \| \cdot \|\mathbf{g}\|} \mathbf{y}, \quad \mathbf{y}\in\mathbb{R}^d.
\end{align}
In fact, $\mathbf{z}$ is still a Gaussian distribution, i.e., $\mathbf{z}\in \mathcal{N}(\mathbf{\mu}, P_{\varphi}\text{diag}(\sigma))$, where $P$ is the projection matrix, with learnable parameters $\varphi$. Thus, the posterior is parameterized by $\theta$ and $\varphi$, i.e., $q_{\phi,\varphi}$. Combining with the equation (\ref{eq:monotonic}), the training objective is  
\begin{align}
    \mathcal{L}(\phi, \theta, \varphi) = &\mathbb{E}_{\mathbf{z}\sim  q_{\phi,\varphi}(\mathbf{z}|\mathbf{v})}[\log p_\theta(\mathbf{y}|\mathbf{z})] \nonumber\\
    + &D_{KL}\left(q_{\phi,\varphi}(\mathbf{z}|\mathbf{v})||p(\mathbf{z})\right)+ R_m,
\end{align}
where $ q_{\phi,\varphi}(\mathbf{z}|\mathbf{v}) = \mathcal{N}(\mathbf{\mu}, P_{\varphi}\text{diag}(\sigma))$. As the latent variable is restricted in space spanned by the vector $\mathbf{g}$, we employ the standard normal distribution on the group interest space as the prior \cite{bahuleyan-etal-2018-variational}. In this case, the closed form of Kullback–Leibler divergence is
\begin{align}
    \sum_{i=1}^d \left(\exp(\sigma_i^\prime)- (1+\sigma_i^\prime + (u_i^\prime)^2\right),
\end{align}
where $\mu_\prime= (\mu_1^\prime, \mu_2^\prime,\dots, \mu_d^\prime)^\top = \text{Proj}_{\mathbf{g}}(\mu)$ and $\sigma^\prime =(\sigma_1^\prime,  \dots, \sigma_d^\prime)^\top=\text{Proj}_{\mathbf{g}}(\sigma)$, respectively

% 先说引入 volume-preserving flow 的必要性
% 然后以一般形式去介绍flow
% 然后说 volume-preserving flow 可以在一定程度上对方差做保护，介绍清楚定理
% 最后详细说到底用了哪一个flow (NICE)
\subsection{Volume-preserving Flow}
The choice of approximate posterior distribution is one of the core problems in variational inference \cite{normalizing_flow}. Most applications of variational inference employ isotropic Gaussians with diagonal covariance for posterior approximations. This restriction has a significant impact on the quality of inferences \cite{normalizing_flow,variational-nmt-flow}. However, such a simple distribution may not be expressive enough to approximate the true posterior distribution. Specifically, the distribution of the user preference can deviate from the Gaussian form, indicating that employing simple distribution families of posterior results in a loose gap between the Evidence Lower Bound (ELBO) and the true marginal likelihood. Inspired by prior works \cite{Introduction-of-flow,NICE,glow}, normalizing flow is a promising method to tackle this problem. The basic idea of normalizing flow is to apply $K$ invertible parametric transformation functions $f_k$ called flows to transform the sample $\mathbf{z}_0 \in \mathbb{R}^d$ of the latent variable. The process is as follows:
\begin{align}
    \mathbf{z}_K = f_{K} \circ f_{K-1} \circ \cdots \circ f_1(\mathbf{z}_0),
\end{align}
where $f_{k}: \mathbb{R}^d \to \mathbb{R}^d$ is an invertible function with Jacobian $\frac{\partial f_k}{\partial \mathbf{z}_{k-1}}$, for $k \in \{1,2,\dots, K\}$. With change-of-variables theorem, the probabilistic density function (PDF) of the random vector $\mathbb{z}_1 $ is given as,
\begin{align}
    p_{\mathbf{z}_1} = p_{\mathbf{z}_0}\left(f_1^{-1}(\mathbf{z}_0)\right) \left|\det \frac{\partial f_1^{-1}}{\partial \mathbf{z}_{0}}\right|.
\end{align}
Following the sequential procedure, we have 
\begin{align}
    \log p_{\mathbf{z}_K} = \log p_{\mathbf{z}_0} - \sum_{k=1}^K \log  \left|\det \frac{\partial f_k}{\partial \mathbf{z}_{k-1}}\right|.
\end{align}
According to the discussion in section \ref{section:Group_Prior}, the variance of the latent variance $\mathbf{z}_K$, regarded as the uncertainty of user interest modeling, the chosen normalizing flow needs to preserve variance of the latent variable sequence $\{\mathbf{z}_1, \mathbf{z}_2, \dots, \mathbf{z}_K\}$. After the invertible transform, the variance of a random variable approximately remains unchanged when the determinant of the Jacobian matrix is equal to 1. The work \cite{book-variance} has shown the following theorem.
\begin{theorem}
    Given a random vector $x$ and a transform $f(\cdot)$, the approximate variance of $f(x)$ is given by 
    \begin{align}
        Var\left[f(x)\right] \approx \left(f^\prime\left|_{x=\mathbb{E}[x]} \right.\right)^2 Var[\mathbf{x} ].
    \end{align}
\end{theorem}
The previous work \cite{article-variance} proves the similar theorme for random vectors. The theorem guides the choice of the normalization flows. As the determinant of the Jacobian matrix of a volume-preserving flow always equals 1, the determinant of the covariance of the latent variable $\mathbf{z}$ approximately remains unchanged after going through a sequence of volume-preserving flows. In this paper, we employ the Non-linear Independent Components Estimation (NICE) developed by \cite{NICE}, which is an instance of a volume-preserving flow. The transformations used are neural networks $f(\cdot)$ with easy to compute inverse $g(\cdot)$ of the form:
\begin{align}
    f(\mathbf{z}) &= \left(\mathbf{z}_A, \mathbf{z}_B + h_\lambda(\mathbf{z}_A) \right)\\
    g(\mathbf{z}^\prime) &= \left(\mathbf{z}^{\prime}_A, \mathbf{z}_B^\prime+ h_\lambda(\mathbf{z}_A) \right),
\end{align}
where $\mathbf{z} = (\mathbf{z}_A, \mathbf{z}_B)$ is a partitioning of the vector $\mathbf{z}$ and $h_\lambda$ is a neural network with parameters $\lambda$. Obviously, the Jacobian matrix of the function $f$ is a lower triangular matrix, resulting in a determinant of 1. Finally, our methods use the output $\mathbf{z}_K$ of flows as the result of user behavior model.

\section{EXPERIMENTS}
% In this section, we present and analyze extensive experiments to show the effectiveness of our proposed method in detail. We show the performance comparison results and present the analysis, including qualitative results, monotonic regularizer visualization, and ablation study. Finally, we give the results of online A/B Testing. 
\subsection{Offline Experiment}
{\bfseries  Datasets.} 1).  The industrial dataset is constructed from the traffic logs of a large-scale recommendation system. We sample a subset of the online logs from 2023-10-05 to 2023-10-16, a total of 12 days, for training and the instances on the next day for testing. This dataset includes substantial details, such as user clicks, user behaviors, user profiles, item profiles, and context features. Besides, we limit the maximal length of the user behavior sequence to 500. 2). The public dataset is Amazon\cite{mcauley2015amazon}. The task is to predict whether a user will write a review for a target item given historical behaviors. Our study follows the settings employed in previous works\cite{DIN,kalman}. 

In detail, we treat the top 25\% users with the most historical interactions as head users and the rest as tail users. We also make the same division on the industrial dataset. Notably, the training set undergoes no processing.

% To validate the effectiveness of our method in behavioral modeling, we selected the widely used Amazon dataset \cite{mcauley2015amazon} as a benchmark for predicting user behavior in click-through rate (CTR) prediction. Our study adheres to the settings employed in previous works\cite{DIN,kalman}. 
% We divide the evaluation by the number of user’s historical interactions. Specifically,  we treat 25\% of the users with the most historical interactions as head users  in public and industrial dataset  and denote the rest as tail users, which is the same as the training phase. 
{\bfseries Metrics.} We adopt AUC as the evaluation metric. AUC represents the probability that a positive sample’s score is higher than a negative one, reflecting a model’s ranking ability \cite{kalman}. AUC is consistent with online performance. For a fair comparison, each model is repeatedly trained and tested 5 times, and the average results are reported.

{\bfseries Baseline.} We evaluate our approach by comparing it to other state-of-the-art user behavior modeling methods. One of these methods, referred to {\bfseries DNN} \cite{covington2016dnn},  treats all user behaviors equally and combines them using sum pooling. An MLP takes the output as the representation of user behaviors. Another method we consider is {\bfseries DIN} \cite{DIN}, aggregating user behaviors attentively using a dedicated local activation unit. Additionally, the Transformer  model \cite{Trans}, namely {\bfseries{Trans}}, is commonly regarded as the state-of-the-art method for capturing sequential signals in behavior sequences. Thus, we compare our proposed method with Trans. 
% The representation is then fed into an MLP. Another method we consider is {\bfseries DIN} \cite{DIN}, which aggregates user behaviors attentively using a dedicated local activation unit. Additionally, we examine the {\bfseries Vanilla Transformer} \cite{Trans}, commonly used for capturing sequential signals in behavior sequences, using settings from previous work \cite{Trans}.

{\bfseries Performance Comparison.}Table \ref{indus_pub_res_exp} depicts a comparison of the test-set AUC in the Click-Through-Rate prediction task. All user behavior modeling methods, such as DIN and Trans, achieve enhancement across all three user groups, when comparing to DNN. The result illustrate the effectiveness of behavior modeling methods. Specially, the Transformer model delivers the best performance, yielding improvements across all three user groups. Thus, we select the Transformer model as the backbone for our method. We note that our proposed method outperforms all other baseline methods, particularly on tail users group, while its performance on head users remains almost the same. For the industrial dataset, we carefully integrate our proposed method with the Transformer model, due to its straightforward implementation. Compared to the Transformer model, our methods improve the AUC by 0.67\% on tail users while maintaining a consistent AUC for head users. Since a 0.1\% improvement in AUC on industrial application is considered significant, our proposed method greatly improves tail users' experiences.

\begin{figure}[!t]
\includegraphics[width=0.5\textwidth]{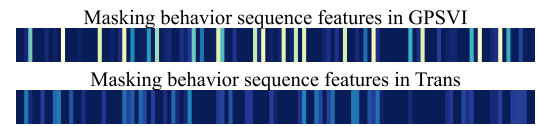}
\caption{Sensitivity of behavior sequence features, visualized by two vectors of size 128.}
  \label{fig:difference_seq}
\end{figure}

{\bfseries Qualitative Results}
We demonstrate GPSVI improved sensitivity to hidden interest of sequences for tail users, visualized by two vectors of size 128 shown in Figure \ref{fig:difference_seq}. Each vector's bins exhibit differences in masking sequence features. The darker the color, the more noticeable the difference. In both cases, we used the well-trained model and visualized the difference in hidden interests (averaged across multiple batches) when the sequence is masked by 0. Clearly, our method enhances sequence modeling among long-tail users,  better utilizing this feature to extract user hidden interests.
% Table \ref{indus_pub_res_exp} depicts a comparison of the test-set AUC of our method in the Click-Through-Rate prediction task. In this table, we first examine the enhancements offered by behavior modeling methods on samples from all users, head users, and tail users. The results show that the Transformer model delivers the best performance, with slight improvements across all three user groups. For the sake of convenience, we have chosen the Transformer model as the backbone for our experiments.We note that our proposed method outperforms all other baseline methods, particularly with tail users, while its functioning with head users remains almost the same. A 0.1\% improvement in AUC on Avito is considered significant. For the industrial dataset, we have carefully pluged our proposed method with the Transformer model, owing to its straightforward implementation. As expected, this approach proves effective. Compared to the Transformer model, our methods improve the AUC by 0.25 on tail users while maintaining a stable AUC on head users.
\begin{table}[!t]
\centering
\small
\caption{Results on industrial  and Amazon dataset for CTR prediction (AUC).The results are averaged over 5 runs. Std $\le$ 0.1\%. }
\label{table1}
\begin{tabular}{l|ccc|ccc}
\hline
\multicolumn{1}{c|}{\multirow{2}{*}{Model}}  & \multicolumn{3}{c|}{Industrial}&\multicolumn{3}{c}{Amazon}\\
\cline{2-7}

\multicolumn{1}{c|}{} &All& Head & Tail &All & Head & Tail \\
\hline

%0.25x-----------------
DNN & 0.7978&0.7982 &0.7853& 0.8346 & 0.8384 &0.8296\\
\cline{1-7}
DIN &0.8005& 0.8009&0.7883&0.8543& 0.8561&0.8517\\
\cline{1-7}
Trans&0.8028&0.8031 &0.7907&0.8666& \textbf{0.8685} &0.8623\\
\cline{1-7}
ours &\textbf{0.8032}&\textbf{ 0.8034}& \textbf{0.7974}& \textbf{0.8682} &0.8684 & \textbf{0.8668}\\
\hline
\end{tabular}
\label{indus_pub_res_exp}
\end{table}

\begin{table}
\centering
\small
\caption{Ablation analysis on proposed method}
\begin{tabular}{lcccccccccc}
\hline
\textbf{Model} &&&&\textbf{All}&&&\textbf{Tail} &&&\textbf{Head} \\
\hline
GPSVI\_wo\_VPF  &&&&0.8023 &&&0.8024 &&&0.7969\\
GPSVI\_wo\_MR   &&&&0.8031 &&&0.8034&&&0.7951\\
\hline
\end{tabular}

\label{tab:ablation}

\end{table}

\begin{figure}[!t]
  \includegraphics[width=0.4\textwidth]{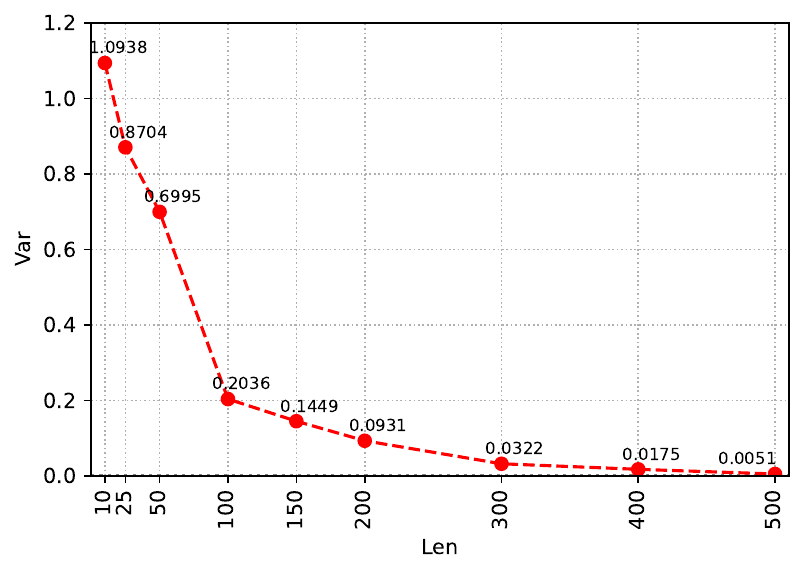}
  \caption{Monotonic regularizer Visualization.}
  \label{fig:seq_len_var}
  % \vspace{}
  %\vspace{-0.5cm}
\end{figure}

{\bfseries Monotonic regularizer Visualization.} In this section, we examine the impact of the monotonicity regularizer on the estimated variance. As depicted in Figure \ref{fig:seq_len_var}, an increase in sequence length corresponds to a decrease in variance. Specifically, the variance drops from approximately 1 when the length $l$ equals 10 to nearly 0.051 when the length reaches 500. This observation demonstrates the significance of monotonic regularity.

{\bfseries Ablation Study.} We conduct an ablation study by removing individual components from the GPSVI model to evaluate their contributions on the industrial dataset. The resulting variants are as follows: {\bfseries GPSVI\_wo\_VPF}, which lacks the volume-preserving flow, and {\bfseries GPSVI\_wo\_MR}, which lacks the monotonic regularizer. Our findings in Table \ref{tab:ablation} reveal that the omission of the monotonic regularizer adversely affects the model's performance on the head users, leading to a 0.1\% decrease in AUC. Conversely, removing the volume-preserving flow does not affect the head users' performance but results in a 0.2\% decrease in AUC for tail users.

% \begin{figure}[!t]
% \includegraphics[width=0.4\textwidth]{tail_heatmap_gpsvi.pdf}
% \caption{Masking behavior sequence features in GPSVI.}
%   \label{fig:seq_len_var}
% \end{figure}

% \begin{figure}[!t]
% \includegraphics[width=0.4\textwidth]{tail_heatmap_base.pdf}
% \caption{Masking behavior sequence features in Trans.}
%   \label{fig:seq_len_var_base}
% \end{figure}

\subsection{ Online A/B tests}
To conduct the online A/B test, we deployed the proposed GPSVI model within a large-scale recommender system. The online A/B test spanned two months. Compared to the baseline model (Transformer), our method resulted in a 0.306\% increase in overall CTR, with a specific improvement of 0.659\% in CTR among tail users.
% Notice that unlike head users, who already present satisfaction with little scope for further enhancement, improving the performance of tail users offers a greater opportunity for significant gains. Our methods have been successfully deployed to serve the main traffic for three months, optimizing user experiences and improving user retention.

% \begin{table}
% \centering
%   \caption{Online A/B test.}
%   \begin{tabular}{lccccccccccccc}
%   \toprule
%      &&&&&&WT&&&&&&&Like\\
%      \hline
%     tail &&&&&& +0.529\% &&&&&&&  1.3\%\\
%     all &&&&&& +0.180\%&&&&&&&0.673\%\\
%   \bottomrule
% \end{tabular}
% \label{tb:ab}
% \end{table}

\section{conclusion}
Attention-based methods have achieved impressive improvement in user behavior modeling, as they emphasize the relevant interactions from rich behaviors. However, the methods encounter challenges capture the preferences of tail users with limited interaction histories. To address the problem, we propose a novel variational inference approach, namely GPSVI, which introduces group preferences as priors to refine latent user interests. Extensive offline experiments and online A/B tests demonstrate that GPSVI consistently improves the performance of tail users. 

\bibliographystyle{acm_format}
% \bibliography{output.bbl}

\begin{thebibliography}{28}

%%% the Web address.

\ifx \showCODEN    \undefined \def \showCODEN     #1{\unskip}     \fi
\ifx \showDOI      \undefined \def \showDOI       #1{#1}\fi
\ifx \showISBNx    \undefined \def \showISBNx     #1{\unskip}     \fi
\ifx \showISBNxiii \undefined \def \showISBNxiii  #1{\unskip}     \fi
\ifx \showISSN     \undefined \def \showISSN      #1{\unskip}     \fi
\ifx \showLCCN     \undefined \def \showLCCN      #1{\unskip}     \fi
\ifx \shownote     \undefined \def \shownote      #1{#1}          \fi
\ifx \showarticletitle \undefined \def \showarticletitle #1{#1}   \fi
\ifx \showURL      \undefined \def \showURL       {\relax}        \fi
% invisible to TeX
\providecommand\bibfield[2]{#2}
\providecommand\bibinfo[2]{#2}
\providecommand\natexlab[1]{#1}
\providecommand\showeprint[2][]{arXiv:#2}

\bibitem[Bahuleyan et~al\mbox{.}(2018)]%
        {bahuleyan-etal-2018-variational}
\bibfield{author}{\bibinfo{person}{Hareesh Bahuleyan}, \bibinfo{person}{Lili Mou}, \bibinfo{person}{Olga Vechtomova}, {and} \bibinfo{person}{Pascal Poupart}.} \bibinfo{year}{2018}\natexlab{}.
\newblock \showarticletitle{Variational Attention for Sequence-to-Sequence Models}. In \bibinfo{booktitle}{\emph{Proceedings of the 27th International Conference on Computational Linguistics}}. \bibinfo{pages}{1672--1682}.
\newblock


\bibitem[Benaroya et~al\mbox{.}(2005)]%
        {book-variance}
\bibfield{author}{\bibinfo{person}{Haym Benaroya}, \bibinfo{person}{Seon~Mi Han}, {and} \bibinfo{person}{Mark Nagurka}.} \bibinfo{year}{2005}\natexlab{}.
\newblock \bibinfo{booktitle}{\emph{Probability Models in Engineering and Science}}.
\newblock \bibinfo{publisher}{CRC Press; 1st edition (June 24, 2005)}.
\newblock


\bibitem[Bowman et~al\mbox{.}(2016)]%
        {bowman-etal-2016-generating}
\bibfield{author}{\bibinfo{person}{Samuel~R. Bowman}, \bibinfo{person}{Luke Vilnis}, \bibinfo{person}{Oriol Vinyals}, \bibinfo{person}{Andrew Dai}, \bibinfo{person}{Rafal Jozefowicz}, {and} \bibinfo{person}{Samy Bengio}.} \bibinfo{year}{2016}\natexlab{}.
\newblock \showarticletitle{Generating Sentences from a Continuous Space}. In \bibinfo{booktitle}{\emph{Proceedings of the 20th {SIGNLL} Conference on Computational Natural Language Learning}}. \bibinfo{pages}{10--21}.
\newblock


\bibitem[Cai et~al\mbox{.}(2023)]%
        {constrained-rl-kuaishou}
\bibfield{author}{\bibinfo{person}{Qingpeng Cai}, \bibinfo{person}{Zhenghai Xue}, \bibinfo{person}{Chi Zhang}, \bibinfo{person}{Wanqi Xue}, \bibinfo{person}{Shuchang Liu}, \bibinfo{person}{Ruohan Zhan}, \bibinfo{person}{Xueliang Wang}, \bibinfo{person}{Tianyou Zuo}, \bibinfo{person}{Wentao Xie}, \bibinfo{person}{Dong Zheng}, \bibinfo{person}{Peng Jiang}, {and} \bibinfo{person}{Kun Gai}.} \bibinfo{year}{2023}\natexlab{}.
\newblock \showarticletitle{Two-Stage Constrained Actor-Critic for Short Video Recommendation}. In \bibinfo{booktitle}{\emph{Proceedings of the ACM Web Conference 2023}} \emph{(\bibinfo{series}{WWW '23})}. \bibinfo{pages}{865–875}.
\newblock


\bibitem[Chen et~al\mbox{.}(2019)]%
        {Trans}
\bibfield{author}{\bibinfo{person}{Qiwei Chen}, \bibinfo{person}{Huan Zhao}, \bibinfo{person}{Wei Li}, \bibinfo{person}{Pipei Huang}, {and} \bibinfo{person}{Wenwu Ou}.} \bibinfo{year}{2019}\natexlab{}.
\newblock \showarticletitle{Behavior sequence transformer for e-commerce recommendation in Alibaba}. In \bibinfo{booktitle}{\emph{Proceedings of the 1st International Workshop on Deep Learning Practice for High-Dimensional Sparse Data}} \emph{(\bibinfo{series}{DLP-KDD '19})}.
\newblock


\bibitem[Covington et~al\mbox{.}(2016)]%
        {covington2016dnn}
\bibfield{author}{\bibinfo{person}{Paul Covington}, \bibinfo{person}{Jay Adams}, {and} \bibinfo{person}{Emre Sargin}.} \bibinfo{year}{2016}\natexlab{}.
\newblock \showarticletitle{Deep neural networks for youtube recommendations}. In \bibinfo{booktitle}{\emph{Proceedings of the 10th ACM conference on recommender systems}}. \bibinfo{pages}{191--198}.
\newblock


\bibitem[Dinh et~al\mbox{.}(2015)]%
        {NICE}
\bibfield{author}{\bibinfo{person}{Laurent Dinh}, \bibinfo{person}{David Krueger}, {and} \bibinfo{person}{Yoshua Bengio}.} \bibinfo{year}{2015}\natexlab{}.
\newblock \showarticletitle{{NICE:} Non-linear Independent Components Estimation}. In \bibinfo{booktitle}{\emph{3rd International Conference on Learning Representations, {ICLR}, Workshop Track Proceedings}}.
\newblock


\bibitem[Du et~al\mbox{.}(2019)]%
        {sequential-meta}
\bibfield{author}{\bibinfo{person}{Zhengxiao Du}, \bibinfo{person}{Xiaowei Wang}, \bibinfo{person}{Hongxia Yang}, \bibinfo{person}{Jingren Zhou}, {and} \bibinfo{person}{Jie Tang}.} \bibinfo{year}{2019}\natexlab{}.
\newblock \showarticletitle{Sequential Scenario-Specific Meta Learner for Online Recommendation}. In \bibinfo{booktitle}{\emph{Proceedings of the 25th ACM SIGKDD International Conference on Knowledge Discovery \& Data Mining}} \emph{(\bibinfo{series}{KDD '19})}. \bibinfo{pages}{2895–2904}.
\newblock


\bibitem[Feng et~al\mbox{.}(2019)]%
        {DSIN}
\bibfield{author}{\bibinfo{person}{Yufei Feng}, \bibinfo{person}{Fuyu Lv}, \bibinfo{person}{Weichen Shen}, \bibinfo{person}{Menghan Wang}, \bibinfo{person}{Fei Sun}, \bibinfo{person}{Yu Zhu}, {and} \bibinfo{person}{Keping Yang}.} \bibinfo{year}{2019}\natexlab{}.
\newblock \showarticletitle{Deep session interest network for click-through rate prediction}. In \bibinfo{booktitle}{\emph{Proceedings of the 28th International Joint Conference on Artificial Intelligence}} \emph{(\bibinfo{series}{IJCAI'19})}. \bibinfo{pages}{2301–2307}.
\newblock


\bibitem[Hou et~al\mbox{.}(2023)]%
        {CIN}
\bibfield{author}{\bibinfo{person}{Xuyang Hou}, \bibinfo{person}{Zhe Wang}, \bibinfo{person}{Qi Liu}, \bibinfo{person}{Tan Qu}, \bibinfo{person}{Jia Cheng}, {and} \bibinfo{person}{Jun Lei}.} \bibinfo{year}{2023}\natexlab{}.
\newblock \showarticletitle{Deep Context Interest Network for Click-Through Rate Prediction}. In \bibinfo{booktitle}{\emph{Proceedings of the 32nd ACM International Conference on Information and Knowledge Management}} \emph{(\bibinfo{series}{CIKM '23})}. \bibinfo{pages}{3948–3952}.
\newblock


\bibitem[Kingma and Dhariwal(2018)]%
        {glow}
\bibfield{author}{\bibinfo{person}{Durk~P Kingma} {and} \bibinfo{person}{Prafulla Dhariwal}.} \bibinfo{year}{2018}\natexlab{}.
\newblock \showarticletitle{Glow: Generative Flow with Invertible 1x1 Convolutions}. In \bibinfo{booktitle}{\emph{Advances in Neural Information Processing Systems}}, Vol.~\bibinfo{volume}{31}.
\newblock


\bibitem[Kingma and Welling(2014)]%
        {Kingma2014}
\bibfield{author}{\bibinfo{person}{Diederik~P. Kingma} {and} \bibinfo{person}{Max Welling}.} \bibinfo{year}{2014}\natexlab{}.
\newblock \showarticletitle{{Auto-Encoding Variational Bayes}}. In \bibinfo{booktitle}{\emph{2nd International Conference on Learning Representations, {ICLR} 2014, Banff, AB, Canada, April 14-16, 2014, Conference Track Proceedings}}.
\newblock


\bibitem[Kobyzev et~al\mbox{.}(2021)]%
        {Introduction-of-flow}
\bibfield{author}{\bibinfo{person}{I. Kobyzev}, \bibinfo{person}{S.~D. Prince}, {and} \bibinfo{person}{M.~A. Brubaker}.} \bibinfo{year}{2021}\natexlab{}.
\newblock \showarticletitle{Normalizing Flows: An Introduction and Review of Current Methods}.
\newblock \bibinfo{journal}{\emph{IEEE Transactions on Pattern Analysis}} \bibinfo{volume}{43}, \bibinfo{number}{11} (\bibinfo{year}{2021}), \bibinfo{pages}{3964--3979}.
\newblock


\bibitem[Liu et~al\mbox{.}(2020)]%
        {kalman}
\bibfield{author}{\bibinfo{person}{Hu Liu}, \bibinfo{person}{Jing Lu}, \bibinfo{person}{Xiwei Zhao}, \bibinfo{person}{Sulong Xu}, \bibinfo{person}{Hao Peng}, \bibinfo{person}{Yutong Liu}, \bibinfo{person}{Zehua Zhang}, \bibinfo{person}{Jian Li}, \bibinfo{person}{Junsheng Jin}, \bibinfo{person}{Yongjun Bao}, {and} \bibinfo{person}{Weipeng Yan}.} \bibinfo{year}{2020}\natexlab{}.
\newblock \showarticletitle{Kalman filtering attention for user behavior modeling in CTR prediction}. In \bibinfo{booktitle}{\emph{Proceedings of the 34th International Conference on Neural Information Processing Systems}} \emph{(\bibinfo{series}{NIPS'20})}.
\newblock


\bibitem[McAuley et~al\mbox{.}(2015)]%
        {mcauley2015amazon}
\bibfield{author}{\bibinfo{person}{Julian McAuley}, \bibinfo{person}{Christopher Targett}, \bibinfo{person}{Qinfeng Shi}, {and} \bibinfo{person}{Anton Van Den~Hengel}.} \bibinfo{year}{2015}\natexlab{}.
\newblock \showarticletitle{Image-based recommendations on styles and substitutes}. In \bibinfo{booktitle}{\emph{Proceedings of the 38th international ACM SIGIR conference on research and development in information retrieval}}. \bibinfo{pages}{43--52}.
\newblock


\bibitem[Mo et~al\mbox{.}(2015)]%
        {image-cold-start}
\bibfield{author}{\bibinfo{person}{Kaixiang Mo}, \bibinfo{person}{Bo Liu}, \bibinfo{person}{Lei Xiao}, \bibinfo{person}{Yong Li}, {and} \bibinfo{person}{Jie Jiang}.} \bibinfo{year}{2015}\natexlab{}.
\newblock \showarticletitle{Image feature learning for cold start problem in display advertising}. In \bibinfo{booktitle}{\emph{Proceedings of the 24th International Conference on Artificial Intelligence}} \emph{(\bibinfo{series}{IJCAI'15})}. \bibinfo{pages}{3728–3734}.
\newblock


\bibitem[Pan et~al\mbox{.}(2019)]%
        {warm-up}
\bibfield{author}{\bibinfo{person}{Feiyang Pan}, \bibinfo{person}{Shuokai Li}, \bibinfo{person}{Xiang Ao}, \bibinfo{person}{Pingzhong Tang}, {and} \bibinfo{person}{Qing He}.} \bibinfo{year}{2019}\natexlab{}.
\newblock \showarticletitle{Warm Up Cold-start Advertisements: Improving CTR Predictions via Learning to Learn ID Embeddings}. In \bibinfo{booktitle}{\emph{Proceedings of the 42nd International ACM SIGIR Conference on Research and Development in Information Retrieval}} \emph{(\bibinfo{series}{SIGIR'19})}. \bibinfo{pages}{695–704}.
\newblock


\bibitem[Pi et~al\mbox{.}(2019)]%
        {MIMN}
\bibfield{author}{\bibinfo{person}{Qi Pi}, \bibinfo{person}{Weijie Bian}, \bibinfo{person}{Guorui Zhou}, \bibinfo{person}{Xiaoqiang Zhu}, {and} \bibinfo{person}{Kun Gai}.} \bibinfo{year}{2019}\natexlab{}.
\newblock \showarticletitle{Practice on Long Sequential User Behavior Modeling for Click-Through Rate Prediction} \emph{(\bibinfo{series}{KDD '19})}. \bibinfo{pages}{2671–2679}.
\newblock


\bibitem[Pi et~al\mbox{.}(2020)]%
        {SIM}
\bibfield{author}{\bibinfo{person}{Qi Pi}, \bibinfo{person}{Guorui Zhou}, \bibinfo{person}{Yujing Zhang}, \bibinfo{person}{Zhe Wang}, \bibinfo{person}{Lejian Ren}, \bibinfo{person}{Ying Fan}, \bibinfo{person}{Xiaoqiang Zhu}, {and} \bibinfo{person}{Kun Gai}.} \bibinfo{year}{2020}\natexlab{}.
\newblock \showarticletitle{Search-based User Interest Modeling with Lifelong Sequential Behavior Data for Click-Through Rate Prediction}. In \bibinfo{booktitle}{\emph{Proceedings of the 29th ACM International Conference on Information \& Knowledge Management}} \emph{(\bibinfo{series}{CIKM '20})}. \bibinfo{pages}{2685–2692}.
\newblock


\bibitem[Rego et~al\mbox{.}(2021)]%
        {article-variance}
\bibfield{author}{\bibinfo{person}{Bruno Rego}, \bibinfo{person}{Dar Weiss}, \bibinfo{person}{Matthew Bersi}, {and} \bibinfo{person}{Jay Humphrey}.} \bibinfo{year}{2021}\natexlab{}.
\newblock \showarticletitle{Uncertainty quantification in subject‐specific estimation of local vessel mechanical properties}.
\newblock \bibinfo{journal}{\emph{International Journal for Numerical Methods in Biomedical Engineering}}  \bibinfo{volume}{37} (\bibinfo{date}{10} \bibinfo{year}{2021}).
\newblock


\bibitem[Rezende and Mohamed(2015)]%
        {normalizing_flow}
\bibfield{author}{\bibinfo{person}{Danilo~Jimenez Rezende} {and} \bibinfo{person}{Shakir Mohamed}.} \bibinfo{year}{2015}\natexlab{}.
\newblock \showarticletitle{Variational inference with normalizing flows}. In \bibinfo{booktitle}{\emph{Proceedings of the 32nd International Conference on International Conference on Machine Learning - Volume 37}} \emph{(\bibinfo{series}{ICML'15})}. \bibinfo{pages}{1530–1538}.
\newblock


\bibitem[Saveski and Mantrach(2014)]%
        {collective-cold}
\bibfield{author}{\bibinfo{person}{Martin Saveski} {and} \bibinfo{person}{Amin Mantrach}.} \bibinfo{year}{2014}\natexlab{}.
\newblock \showarticletitle{Item cold-start recommendations: learning local collective embeddings}. In \bibinfo{booktitle}{\emph{Proceedings of the 8th ACM Conference on Recommender Systems}} \emph{(\bibinfo{series}{RecSys '14})}. \bibinfo{pages}{89–96}.
\newblock


\bibitem[Setiawan et~al\mbox{.}(2020)]%
        {variational-nmt-flow}
\bibfield{author}{\bibinfo{person}{Hendra Setiawan}, \bibinfo{person}{Matthias Sperber}, \bibinfo{person}{Udhyakumar Nallasamy}, {and} \bibinfo{person}{Matthias Paulik}.} \bibinfo{year}{2020}\natexlab{}.
\newblock \showarticletitle{Variational Neural Machine Translation with Normalizing Flows}. In \bibinfo{booktitle}{\emph{Proceedings of the 58th Annual Meeting of the Association for Computational Linguistics}}. \bibinfo{pages}{7771--7777}.
\newblock


\bibitem[Volkovs et~al\mbox{.}(2017)]%
        {NIPS2017_dbd22ba3}
\bibfield{author}{\bibinfo{person}{Maksims Volkovs}, \bibinfo{person}{Guangwei Yu}, {and} \bibinfo{person}{Tomi Poutanen}.} \bibinfo{year}{2017}\natexlab{}.
\newblock \showarticletitle{DropoutNet: Addressing Cold Start in Recommender Systems}. In \bibinfo{booktitle}{\emph{Advances in Neural Information Processing Systems}} \emph{(\bibinfo{series}{NIPS '14})}.
\newblock


\bibitem[Xu et~al\mbox{.}(2022)]%
        {velf}
\bibfield{author}{\bibinfo{person}{Xiaoxiao Xu}, \bibinfo{person}{Chen Yang}, \bibinfo{person}{Qian Yu}, \bibinfo{person}{Zhiwei Fang}, \bibinfo{person}{Jiaxing Wang}, \bibinfo{person}{Chaosheng Fan}, \bibinfo{person}{Yang He}, \bibinfo{person}{Changping Peng}, \bibinfo{person}{Zhangang Lin}, {and} \bibinfo{person}{Jingping Shao}.} \bibinfo{year}{2022}\natexlab{}.
\newblock \showarticletitle{Alleviating Cold-start Problem in CTR Prediction with A Variational Embedding Learning Framework}. In \bibinfo{booktitle}{\emph{Proceedings of the ACM Web Conference 2022}} \emph{(\bibinfo{series}{WWW '22})}. \bibinfo{pages}{27–35}.
\newblock


\bibitem[Yin et~al\mbox{.}(2020)]%
        {learningtrans}
\bibfield{author}{\bibinfo{person}{Jianwen Yin}, \bibinfo{person}{Chenghao Liu}, \bibinfo{person}{Weiqing Wang}, \bibinfo{person}{Jianling Sun}, {and} \bibinfo{person}{Steven~C.H. Hoi}.} \bibinfo{year}{2020}\natexlab{}.
\newblock \showarticletitle{Learning Transferrable Parameters for Long-tailed Sequential User Behavior Modeling}. In \bibinfo{booktitle}{\emph{Proceedings of the 26th ACM SIGKDD International Conference on Knowledge Discovery \& Data Mining}} \emph{(\bibinfo{series}{KDD '20})}. \bibinfo{pages}{359–367}.
\newblock


\bibitem[Zhou et~al\mbox{.}(2019)]%
        {DIEN}
\bibfield{author}{\bibinfo{person}{Guorui Zhou}, \bibinfo{person}{Na Mou}, \bibinfo{person}{Ying Fan}, \bibinfo{person}{Qi Pi}, \bibinfo{person}{Weijie Bian}, \bibinfo{person}{Chang Zhou}, \bibinfo{person}{Xiaoqiang Zhu}, {and} \bibinfo{person}{Kun Gai}.} \bibinfo{year}{2019}\natexlab{}.
\newblock \showarticletitle{Deep interest evolution network for click-through rate prediction}. In \bibinfo{booktitle}{\emph{Proceedings of the Thirty-Third AAAI Conference on Artificial Intelligence and Thirty-First Innovative Applications of Artificial Intelligence Conference and Ninth AAAI Symposium on Educational Advances in Artificial Intelligence}} \emph{(\bibinfo{series}{AAAI '19})}.
\newblock


\bibitem[Zhou et~al\mbox{.}(2018)]%
        {DIN}
\bibfield{author}{\bibinfo{person}{Guorui Zhou}, \bibinfo{person}{Xiaoqiang Zhu}, \bibinfo{person}{Chenru Song}, \bibinfo{person}{Ying Fan}, \bibinfo{person}{Han Zhu}, \bibinfo{person}{Xiao Ma}, \bibinfo{person}{Yanghui Yan}, \bibinfo{person}{Junqi Jin}, \bibinfo{person}{Han Li}, {and} \bibinfo{person}{Kun Gai}.} \bibinfo{year}{2018}\natexlab{}.
\newblock \showarticletitle{Deep Interest Network for Click-Through Rate Prediction}. In \bibinfo{booktitle}{\emph{Proceedings of the 24th ACM SIGKDD International Conference on Knowledge Discovery \& Data Mining}} \emph{(\bibinfo{series}{KDD '18})}. \bibinfo{pages}{1059–1068}.
\newblock


\end{thebibliography}

\appendix









\end{document}